\documentclass[%
aps, %
pra, 
preprint, onecolumn, superscriptaddress, 11pt, 
citeautoscript, %
a4paper, %
eqsecnum, 
amsfonts, amsmath, amssymb]{revtex4-1}
\usepackage[sort&compress]{natbib}
\usepackage[varg]{txfonts}
\usepackage{anyfontsize}
\usepackage{ascmac}
\usepackage{mathrsfs}
\usepackage{bm}
\usepackage{color}
\usepackage{graphicx}
\usepackage{bm}
\usepackage{braket}
\usepackage{mathrsfs}
\usepackage[normalem]{ulem}
\usepackage{hhline}
\usepackage{hyperref}
\hypersetup{%
	unicode=false, %
	bookmarksnumbered=true, %
	bookmarkstype=toc, %
	breaklinks=true, %
	colorlinks=true, %
	pdfborder={0 0 1}, %
	linkcolor=blue, %
	citecolor=blue, %
	urlcolor=blue, %
	pdftoolbar=true, %
	pdfmenubar=true, %
	pdffitwindow=false, %
	pdfstartview={FitH} %
}
\renewcommand{\Im}{\mathrm{Im}\,}
\graphicspath{{./}}
\begin{document}
%
%
\title{Intrinsic and Extrinsic Spin Hall Effects of Dirac Electrons}
\date{\today}
\author{Takaaki Fukazawa}
\author{Hiroshi Kohno}
\affiliation{Department of Physics, Nagoya University, Furo-cho, Chikusa-ku, Nagoya 464-8602, Japan}
\author{Junji Fujimoto}
\email[Corresponding author: ]{fujimoto.junji.s8@kyoto-u.ac.jp}
\affiliation{Institute for Chemical Research, Kyoto University, Uji, Kyoto 611-0011, Japan}
\affiliation{RIKEN Center for Emergent Matter Science, Wako, Saitama 351-0198, Japan}
\begin{abstract}
 We investigate the spin Hall effect (SHE) of electrons described by the Dirac equation, which is used as an effective model near the $L$-points in bismuth.
 By considering short-range nonmagnetic impurities, we calculate the extrinsic as well as intrinsic contributions on an equal footing. 
 The vertex corrections are taken into account within the ladder type and the so-called skew-scattering type.
 The intrinsic SHE which we obtain is consistent with that of Fuseya {\it et al.} [J. Phys. Soc. Jpn. {\bf 81}, 93704 (2012)].
 It is found that the extrinsic contribution dominates the intrinsic one when the system is metallic.
 The extrinsic SHE due to the skew scattering is proportional to $\Delta / n_{\rm i} u$, where $2\Delta$ is the band gap, $n_{\rm i}$ is the impurity concentration, and $u$ is the strength of the impurity potential.
\end{abstract}
\maketitle
\section{\label{sec:introduction}Introduction}
 The spin Hall effect (SHE)~\cite{Sinova2015} is one of fascinating phenomena, which allows us to convert electric currents into pure spin currents directly.
 It is the effect that when applying the uniform electric field in the $\hat{x}$-direction, $E_x$, the spin current which has the $\hat{z}$-directed spin polarization flows in the $\hat{y}$-direction: $J^{z}_{s, y} = \sigma^{z}_{\mathrm{s}, yx} E_x$, where $\sigma^{z}_{\mathrm{s}, yx}$ is the spin Hall conductivity.
 The SHE as well as other spin-dependent Hall effects, such as the anomalous Hall effect~\cite{Nagaosa2010}, originates from the spin-orbit coupling~(SOC) and quantum coherent band mixing by the external electric field and/or the impurity potential.
 Three distinct contributions to the SHE are known: intrinsic~\cite{Murakami2003,Sinova2004}, side-jump~\cite{Berger1970} and skew-scattering~\cite{Smit1955,Smit1958}.
 The skew-scattering contribution is defined as a component proportional to $\tau$ (or $n_{\rm i}^{-1}$), where $\tau$ is the lifetime of electrons and $n_{\rm i}$ is  
the impurity concentration.
 The intrinsic and side-jump contributions are independent of $n_{\rm i}$.
 The intrinsic contribution is defined as the one without any impurity potentials, and the side-jump contribution is distinguished by subtracting the intrinsic one from the $n_{\rm i}$-independent components.
 The extrinsic (skew-scattering and side-jump) contributions are theoretically evaluated as the self energy and the vertex corrections~(VCs) due to the impurity potentials.

 A lot of quantum mechanical phenomena were first discovered in bismuth, such as the Shubnikov-de Haas oscillation~\cite{Shubnikov1930} and de Haas-van Alphen effect~\cite{deHaas1930}.
 It was shown that the conduction/valence electrons near the $L$-points in bismuth are effectively described by the Dirac-type Hamiltonian~\cite{Cohen1960,Wolff1964,Fuseya2015}.
 The large diamagnetism of bismuth~\cite{Wehrli1968} was explained by the interband effect of the magnetic field assisted by the large SOC based on the effective Hamiltonian~\cite{Fukuyama1970}.
 Considering the strong SOC in bismuth, large SHE was theoretically expected by Fuseya {\it et al.}~\cite{Fuseya2012}.
 They calculated the intrinsic contribution to the SHE, and found a simple relation to the orbital susceptibility when the chemical potential lies in the gap ($|\mu / \Delta| < 1$, $\mu$ is the chemical potential and $2 \Delta$ is the band gap)~\cite{Fuseya2012}.
 
 The extrinsic contributions are known to give rise to important effects.
 Particularly for the two-dimensional~(2D) Rashba system with $\delta$-function type impurity potentials, the side-jump contribution cancels out the intrinsic one~\cite{Inoue2004}.
 The skew-scattering contribution becomes dominant for the cleaner system in general, while it is absent in the 2D Rashba system~\cite{Inoue2006}.
 In this paper, we take into account the $\delta$-function type of non-magnetic impurity potentials for the Dirac Hamiltonian and calculate the extrinsic contributions to the SHE as well as the intrinsic one on an equal footing.
 The vertex corrections are considered within the ladder type and the so-called skew-scattering type.
 The results of the intrinsic SHE agree with that of Fuseya {\it et al.}~\cite{Fuseya2012}.
 We find that the extrinsic SHE dominates over the intrinsic one when the system is metallic ($|\mu / \Delta| > 1$), while the intrinsic SHE has a peak when $|\mu / \Delta| < 1$.
 The skew-scattering contribution is finite when $|\mu / \Delta| > 1$ and gives rise to a significant contribution to the SHE ($\propto \Delta / n_{\rm i} u \equiv \eta^{-1}$), if $\eta \lesssim 0.3$, where $u$ is the strength of the impurity potential.

\section{\label{sec:Model_and_Green_functions}Model and Green functions}
 Following Ref.~\cite{Fuseya2012}, we consider the effective (isotropic) Dirac Hamiltonian,
\begin{align}
\mathcal{H}_{\mathrm{D}}
	& =
	\begin{pmatrix}
		\Delta
	&	i \hbar v \bm{k} \cdot \bm{\sigma}
	\\	- i \hbar v \bm{k} \cdot \bm{\sigma}
	&	- \Delta
	\end{pmatrix}
	= - \hbar v \rho_2 \bm{k} \cdot \bm{\sigma} + \Delta \rho_3
\label{eq:Dirac_Hamiltonian}
,\end{align}
where $\pm v$ is the velocity of the Dirac electrons, $\bm{\sigma} = (\sigma^{x}, \sigma^{y}, \sigma^{z})$ are the Pauli matrices in spin space, and $\rho_i$, $i=1,2,3$ are the Pauli matrices in particle-hole space.
 We also use $\rho_0$ and $\sigma^0$ as the unit matrices when we emphasize them.
 The eigen energies of this Hamiltonian are given by $\pm \varepsilon_k = \pm \sqrt{\hbar^2 v^2 k^2 + \Delta^2}$.
\begin{align}
v_i
	& = \frac{1}{\hbar} \frac{ \partial \mathcal{H}_{\mathrm{D}} }{ \partial k_i }
	= - v \rho_2 \sigma^i
\qquad (i = x, y, z)
\label{eq:particle_velocity}
\end{align}
is the velocity operator, and the velocity of the spin current with spin component $\alpha$ is given by
\begin{align}
v_{\mathrm{s}, i}^{\alpha}
	& = \frac{1}{2 \mu_{\mathrm{B}}} \bigl[ v_i, \mu_{\mathrm{s}, \alpha} \bigr]_{+}
\qquad (i, \alpha = x, y, z)
\\	&
	 = - \frac{g^* v}{2} \epsilon_{i \alpha j} \rho_1 \sigma^j
\label{eq:spin_velocity}
,\end{align}
where $[A, B]_{+} = AB + BA$, and $\bm{\mu}_{\mathrm{s}} = - (g^* \mu_{\mathrm{B}} / 2) \rho_3 \bm{\sigma}$ is the spin magnetic moment with $g^{*} = 2 m v^2 / \Delta$ being the effective $g$-factor and $\mu_{\mathrm{B}}$ the Bohr magneton.
 To be precise, $\bm{v}_{\mathrm{s}}^{\alpha}$ represents the velocity of the spin-magnetic-moment current, but we call it that of the spin current in this paper.
 Hereafter, we put $\hbar = 1$.

 It is not obvious how the impurity potential is expressed in the basis of the Dirac Hamiltonian as will be commented in the end of this paper.
For the present, we assume that it is proportional to a unit matrix,
\begin{align}
V_{\rm imp} (\bm{r})
	& = u\,\rho_0 \sigma^0 \sum_{i=1}^{N_{\rm i}} \delta(\bm{r} - \bm{R}_j)
\label{eq:impurity_Hamiltonian}
,\end{align}
and treat it within the Born approximation for the self energy.
 (The justification of this approximation will be commented later in this section.)
 Here, $N_{\rm i}$ is the number of the impurities, $u$ is the strength of the impurity potential, and $\bm{R}_j$ is the position of the $j$-th impurity.
 By taking the average on the positions of the impurities~\cite{Kohn1957,Mahan2000}, the retarded self energy is given by
\begin{align}
\varSigma^{\rm R} (\epsilon)
	& = \frac{n_{\rm i} u^2}{\Omega} \sum_{\bm{k}} G^{(0)}_{\bm{k}} (\epsilon + i 0)
\label{eq:retarded_self_energy}
,\end{align}
where $n_{\rm i} = N_{\rm i} / \Omega$ with $\Omega$ being the volume of the system, and the bare Green function is defined by $G^{(0)}_{\bm{k}} (\epsilon + i 0) = (\epsilon + \mu - \mathcal{H}_{\rm D} + i 0)^{-1}$.
 The imaginary part is evaluated as
\begin{align}
\Im \varSigma^{\rm R} (\epsilon)
	& = - \gamma_0 (\epsilon+\mu) - \gamma_3 (\epsilon+\mu) \rho_3
\label{eq:Im_self-energy}
\end{align}
with
\begin{subequations}
\begin{align}
\gamma_0 (\epsilon)
	& = \frac{\pi}{2} n_{\rm i} u^2 \nu (\epsilon)
, \\
\gamma_3 (\epsilon)
	& = \frac{\pi}{2} n_{\rm i} u^2 \frac{\Delta}{\epsilon} \nu (\epsilon)
,\end{align}
\label{eq:damping-constants}%
\end{subequations}
where $\nu (\epsilon)$ is the density of states~(DOS),
\begin{align}
\nu (\epsilon)
	& = \frac{1}{\Omega} \sum_{\bm{k}, \eta = \pm} \delta (\epsilon - \eta \varepsilon_k)
	= \frac{|\epsilon|}{2 \pi^2 v^3} \sqrt{\epsilon^2 - \Delta^2} \sum_{\eta} \Theta ( \eta \epsilon - \Delta )
\label{eq:DOS}
\end{align}
with $\Theta (x)$ being the Heaviside step function,
\begin{align*}
\Theta (x)
	& = \int_{-\infty}^{x} \mathrm{d} t \, \delta (t)
	= \left\{
	\begin{array}{cc}
		1
	&	(x > 0)
	, \\	0
	&	(x < 0)
	.\end{array}
\right.
\end{align*}
 From these, the retarded Green function including the self energy is obtained as
\begin{align}
G^{\mathrm{R}}_{\bm{k}} (\epsilon)
	& = \frac{1}{ D^{\mathrm{R}}_{\bm{k}} (\epsilon+\mu) }
		\Bigl( g^{\mathrm{R}}_0 (\epsilon+\mu) + \rho_2 \bm{g}^{\mathrm{R}}_2 (\bm{k}) \cdot \bm{\sigma} + g^{\mathrm{R}}_3 (\epsilon+\mu) \rho_3 \Bigr)
\label{eq:retarded_G_def}
,\end{align}
where
\begin{subequations}
\begin{align}
D^{\mathrm{R}}_{\bm{k}} (\epsilon)
	& = ( \epsilon + i \gamma_0 (\epsilon) )^2
	- v^2 k^2
	- (\Delta - i \gamma_3 (\epsilon))^2
\label{eq:DR_k}
, \\
g^{\mathrm{R}}_0 (\epsilon)
	& = \epsilon + i \gamma_0 (\epsilon)
, \\
\bm{g}^{\mathrm{R}}_2 (\bm{k})
	& = - v \bm{k}
, \\
g^{\mathrm{R}}_3 (\epsilon)
	& = \Delta - i \gamma_3 (\epsilon)
.\end{align}
\end{subequations}
We dropped the real part of the self energy since they are just $\mathcal{O} (n_{\rm i})$ and can be absorbed into $\mu$ and $\Delta$.

 To include skew scattering in the above formalism, we first consider the self-consistent $T$-matrix approximation, and then restrict to the case $n_{\rm i} \ll 1$ and $\pi u \nu (\mu) \ll 1$.
Then, the self energy reduces to the one in the Born approximation [Eq.~(\ref{eq:retarded_self_energy})].

\section{\label{sec:spin_Hall_conductivity}Calculation of spin Hall conductivity}
 Now we calculate the spin Hall conductivity $\sigma^{z}_{\mathrm{s}, yx}$.
 According to the Kubo formula, $\sigma^{z}_{\mathrm{s}, yx}$ is calculated from the $\omega$-linear term of the correlation function between the spin current and electric current, where $\omega$ is the frequency of the applied electric field.
 In this paper, we assume that the impurities are dilute and evaluate the spin Hall conductivity in the leading order of $n_{\rm i}$.
 After a straightforward calculation
, we obtain
\begin{align}
\sigma^{z}_{\mathrm{s}, yx}
	& = \frac{- e g^* v^2}{\pi} ( K^{\mathrm{b}}_{\mathrm{surf}} + K^{\mathrm{VC}}_{\mathrm{surf}} + K^{}_{\mathrm{sea}} )
\label{eq:spin_Hall_conductivity}
,\end{align}
where
\begin{align}
K^{\mathrm{b}}_{\mathrm{surf}}
	& = \frac{1}{4 \Omega} \int_{-\infty}^{\infty} \mathrm{d} \epsilon \left( - \frac{\partial f}{\partial \epsilon} \right)
			\sum_{\bm{k}} \mathrm{tr} \left[
				\rho_1 \sigma^x G^{\mathrm{R}}_{\bm{k}} (\epsilon) \rho_2 \sigma^x G^{\mathrm{A}}_{\bm{k}} (\epsilon)
			\right]
\label{eq:sHc_bare_def}
\end{align}
is the bare bubble contribution from the states near the Fermi level, $K_{\mathrm{surf}}^{\mathrm{VC}}$ is the vertex correction to be considered later, and
\begin{align}
K^{}_{\mathrm{sea}}
	& = \frac{1}{8 \Omega} \int_{-\infty}^{\infty} \mathrm{d} \epsilon f (\epsilon)
			\sum_{\bm{k}} \mathrm{tr} \left[
				\rho_1 \sigma^x G^{\mathrm{R}}_{\bm{k}} (\epsilon) \rho_2 \sigma^x \Bigl(\partial_{\epsilon} G^{\mathrm{R}}_{\bm{k}} (\epsilon) \Bigr)
				- \rho_1 \sigma^x \Bigl(\partial_{\epsilon} G^{\mathrm{R}}_{\bm{k}} (\epsilon) \Bigr) \rho_2 \sigma^x G^{\mathrm{R}}_{\bm{k}} (\epsilon)
				- \Bigl( \mathrm{R} \leftrightarrow \mathrm{A} \Bigr)
				 \right]
\label{eq:sHc_sea_def}
\end{align}
is the bare bubble contribution from the states below the Fermi level.
 Here, $f (\epsilon) = (e^{\epsilon / k_{\mathrm{B}} T} + 1)^{-1}$.

 It is convenient to first look at the following quantities,
\begin{subequations}
\begin{align}
\frac{n_{\rm i} u^2}{\Omega} \sum_{\bm{k}}
G^{\mathrm{R}}_{\bm{k}} (\epsilon) \rho_1 \sigma^x G^{\mathrm{A}}_{\bm{k}} (\epsilon)
	& = 2 U (\epsilon+\mu) \rho_1 \sigma^x
		+ V (\epsilon+\mu) \rho_2 \sigma^x
		+ \mathcal{O} (n_{\rm i}^2)
\label{eq:GR_rho1_GA}
, \\
\frac{n_{\rm i} u^2}{\Omega} \sum_{\bm{k}}
G^{\mathrm{R}}_{\bm{k}} (\epsilon) \rho_2 \sigma^x G^{\mathrm{A}}_{\bm{k}} (\epsilon)
	& = - V (\epsilon+\mu) \rho_1 \sigma^x
		+ U (\epsilon+\mu) \rho_2 \sigma^x
		+ \mathcal{O} (n_{\rm i}^2)
\label{eq:GR_rho2_GA}
,\end{align}
\label{eqs:GR_rho1_GA-GR_rho2_GA}%
\end{subequations}
where $U (\epsilon)$ and $V (\epsilon)$ are $\mathcal{O} (n_{\rm i}^0)$- and $\mathcal{O} (n_{\rm i})$-terms, respectively, given by
\begin{align}
U (\epsilon)
	& = \frac{1}{4} \frac{n_{\rm i} u^2}{\Omega} \sum_{\bm{k}} \mathrm{tr} \Bigl[
	G^{\mathrm{R}}_{\bm{k}} (\epsilon-\mu) \rho_2 \sigma^x G^{\mathrm{A}}_{\bm{k}} (\epsilon-\mu) \rho_2 \sigma^x
	\Bigr]
\label{eq:U_def}
\\ &
	= \frac{1}{3} \frac{ \epsilon^2 - \Delta^2 }{ \epsilon^2 + \Delta^2 }
	+ \mathcal{O} (\gamma^2)
\label{eq:longitudinal_conductivity}
, \\
V (\epsilon)
	& = \frac{1}{4} \frac{n_{\rm i} u^2}{\Omega} \sum_{\bm{k}} \mathrm{tr} \Bigl[
	G^{\mathrm{R}}_{\bm{k}} (\epsilon-\mu) \rho_1 \sigma^x G^{\mathrm{A}}_{\bm{k}} (\epsilon-\mu) \rho_2 \sigma^x
	\Bigr]
\label{eq:V_def}
\\ &
	= \frac{\pi}{2 |\epsilon|} n_{\rm i} u^2 \frac{ \Delta \gamma_0 (\epsilon) + \epsilon \gamma_3 (\epsilon) }{ | \epsilon \gamma_0 (\epsilon) + \Delta \gamma_3 (\epsilon) | } \nu (\epsilon)
	+ \mathcal{O} (\gamma^3)
\label{eq:limit_line}
\\ &
	= n_{\rm i} u^2 \frac{ \pi \Delta }{ \epsilon^2 + \Delta^2 } \nu (\epsilon)
	+ \mathcal{O} (\gamma^3)
\notag
.\end{align}
 In carrying out the $\bm{k}$-integrals in Eqs.~(\ref{eq:U_def}) and (\ref{eq:V_def}), we expressed as $D^{\mathrm{R}}_{\bm{k}} (\epsilon) = D' + i D''$ [Eq.~(\ref{eq:DR_k})] with
\begin{subequations}
\begin{align}
D'
	& = (\epsilon - \varepsilon_k) (\epsilon + \varepsilon_k)
		+ \mathcal{O} (\gamma^2)
, \\
D''
	& = 2 ( \epsilon \gamma_0 (\epsilon) + \Delta \gamma_3 (\epsilon) )
,\end{align}%
\end{subequations}
and used the following approximations,
\begin{align}
\frac{1}{|D^{\mathrm{R}}_{\bm{k}} (\epsilon)|^2}
	& \simeq \frac{\pi}{|D''|} \delta (D')
\notag \\	&
	\simeq \frac{\pi}{4} \frac{1}{|\epsilon \gamma_0 (\epsilon) + \Delta \gamma_3 (\epsilon)|} \frac{1}{|\epsilon|}
		\sum_{\eta = \pm} \delta ( \epsilon - \eta \varepsilon_k )
.\end{align}
Here, by assuming small $n_{\rm i}$, we approximated $D''/( (D')^2 + (D'')^2)$ by the $\delta$-function in the first line, and dropped $\gamma_{0}^2$ and $\gamma_3^2$ in the second line.
 Using Eq.~(\ref{eq:GR_rho2_GA}), we obtain
\begin{align}
K_{\mathrm{surf}}^{\mathrm{b}}
	& = - \int_{-\infty}^{\infty} \mathrm{d} \epsilon \left( - \frac{\partial f_{\mathrm{FD}}}{\partial \epsilon} \right) \frac{V (\epsilon)}{n_{\rm i} u^2}
\notag \\
	& = - \int_{-\infty}^{\infty} \mathrm{d} \epsilon \left( - \frac{\partial f_{\mathrm{FD}}}{\partial \epsilon} \right)
		\frac{ \pi \Delta }{ \epsilon^2 + \Delta^2 } \nu (\epsilon)
\label{eq:sHc_bare}
,\end{align}
where $f_{\mathrm{FD}} (\epsilon) =  (e^{(\epsilon - \mu)/ k_{\mathrm{B}} T} + 1)^{-1}$ is the Fermi-Dirac distribution function.

 Next, we calculate $K_{\mathrm{sea}}^{}$.
 We can put $n_{\rm i} \to 0$ since it has no singularities.
 This means that $K_{\mathrm{sea}}^{} (n_{\rm i} \to 0)$ contains an intrinsic contribution only.
 Hence, we write $K_{\mathrm{sea}}^{\rm (int)} \equiv K_{\mathrm{sea}}^{} (n_{\rm i} \to 0)$.
 The trace part in Eq.~(\ref{eq:sHc_sea_def}) is calculated as
\begin{align}
\mathrm{tr} [ \,\cdots ]
	& = - 8 i \Delta \left( \frac{1}{ (D' + i D'')^2 } - \frac{1}{ (D' - i D'')^2 } \right)
\notag	\\
	& = - 8 i \Delta \frac{1}{\partial_{\epsilon} D'} \frac{\partial}{\partial \epsilon} \left( \frac{1}{ D' + i D'' } - \frac{1}{ D' - i D'' } \right)
\notag	\\	
	& = - 16 \pi \Delta \frac{\mathrm{sgn} (D'')}{\partial_{\epsilon} D'} \frac{\partial}{\partial \epsilon} \delta (D')
\qquad (n_{\rm i} \to 0)
\label{eq:DR2-DA2}
.\end{align}
 Integrating by parts and using $\delta (D') = \sum_{\eta = \pm} \delta ( \epsilon - \eta \varepsilon_k ) / |\epsilon|$, $\partial_{\epsilon} D' = 2 \epsilon$, and $\mathrm{sgn} (D'') = \mathrm{sgn} (\epsilon)$,
then we obtain
\begin{align}
K^{\rm (int)}_{\mathrm{sea}}
	& = \int_{-\infty}^{\infty} \mathrm{d} \epsilon \left( - \frac{\partial f_{\mathrm{FD}}}{\partial \epsilon} \right)
	\frac{\pi \Delta}{2 \epsilon^2} \nu (\epsilon)
	+ \int_{-\infty}^{\infty} \mathrm{d} \epsilon f_{\mathrm{FD}} (\epsilon)
		\frac{\pi \Delta}{2 \epsilon^3} \nu (\epsilon)
.\end{align}
 We see that $K^{\rm (int)}_{\mathrm{sea}}$ also contains the contribution from the states near the Fermi level.

 Finally, we estimate $K_{\mathrm{surf}}^{\mathrm{VC}}$, which consists of terms with the vertex corrections of the ladder type and the so-called skew-scattering type, denoted as $K_{\mathrm{surf}}^{\mathrm{ld}}$ and $K_{\mathrm{surf}}^{\mathrm{sk}}$, respectively.
 They are given by
\begin{align}
K_{\mathrm{surf}}^{\mathrm{ld}}
	& = \frac{1}{4 \Omega} \int_{-\infty}^{\infty} \mathrm{d} \epsilon \left( - \frac{\partial f}{\partial \epsilon} \right)
			\sum_{\bm{k}} \mathrm{tr} \left[
				\rho_1 \sigma^x
				G^{\mathrm{R}}_{\bm{k}} \tilde{\Lambda}_{2,x} G^{\mathrm{A}}_{\bm{k}}
			\right]
, \\
K_{\mathrm{surf}}^{\mathrm{sk}}
	& = \frac{ n_{\rm i} u^3 }{4 \Omega^3} \int_{-\infty}^{\infty} \mathrm{d} \epsilon \left( - \frac{\partial f}{\partial \epsilon} \right)
			\sum_{\bm{k}, \bm{k}', \bm{k}''} \mathrm{tr} \left[
				\Lambda_{1,x}^{*} G^{\mathrm{R}}_{\bm{k}}
				G^{\mathrm{R}}_{\bm{k}'} \Lambda_{2,x} G^{\mathrm{A}}_{\bm{k}'}
				G^{\mathrm{A}}_{\bm{k}''}
				G^{\mathrm{A}}_{\bm{k}}
				+ \Lambda_{1,x}^{*} G^{\mathrm{R}}_{\bm{k}}
				G^{\mathrm{R}}_{\bm{k}''}
				G^{\mathrm{R}}_{\bm{k}'} \Lambda_{2,x} G^{\mathrm{A}}_{\bm{k}'}
				G^{\mathrm{A}}_{\bm{k}}
			\right]
,\end{align}
where $G^{\mathrm{R/A}}_{\bm{k}} = G^{\mathrm{R/A}}_{\bm{k}} (\epsilon)$, and $\Lambda_{1,x} = \Lambda_{1,x} (\epsilon)$ and $\Lambda_{2,x} = \Lambda_{2,x} (\epsilon)$ are proportional to the full velocities of the spin current [Eq.~(\ref{eq:spin_velocity})] and the particle current [Eq.~(\ref{eq:particle_velocity})],
\begin{align}
\Lambda_{1,x}
	& = \rho_1 \sigma^x
		+ \frac{n_{\rm i} u^2}{\Omega} \sum_{\bm{k}} G^{\mathrm{R}}_{\bm{k}} \Lambda_{1,x} G^{\mathrm{A}}_{\bm{k}}
\label{eq:Lambda_1x_RA}
, \\
\Lambda_{2,x}
	& = \rho_2 \sigma^x
		+ \frac{n_{\rm i} u^2}{\Omega} \sum_{\bm{k}} G^{\mathrm{R}}_{\bm{k}} \Lambda_{2,x} G^{\mathrm{A}}_{\bm{k}}
\label{eq:Lambda_2x_RA}
,\end{align}
$\tilde{\Lambda}_{2,x} = \Lambda_{2,x} - \rho_2 \sigma^x$ is VC, and $\Lambda_{1,x}^{*}$ is defined by interchanging $\mathrm{R}$ and $\mathrm{A}$ in Eq.~(\ref{eq:Lambda_1x_RA}).
 Using Eqs.~(\ref{eq:GR_rho1_GA}) and (\ref{eq:GR_rho2_GA}), these VCs up to $\mathcal{O} (n_{\rm i})$ are calculated as
\begin{align}
\begin{pmatrix}
	\Lambda_{1,x}
\\	\Lambda_{2,x}
\end{pmatrix}
	& =
		\sum_{n=0}^{\infty}
		\begin{pmatrix}
			2 U
		&	V
		\\	- V
		&	U
		\end{pmatrix}^n
	\begin{pmatrix}
		\rho_1 \sigma^x
	\\	\rho_2 \sigma^x
	\end{pmatrix}
	+ \mathcal{O} ( n_{\rm i}^2 )
\notag
	 =
	\left( \hat{A} + \hat{A} \hat{B} \hat{A} \right)
	\begin{pmatrix}
		\rho_1 \sigma^x
	\\	\rho_2 \sigma^x
	\end{pmatrix}
	+ \mathcal{O} ( n_{\rm i}^2 )
,\end{align}
where $U = U (\epsilon + \mu)$, $V = V (\epsilon + \mu)$, and
\begin{align}
\hat{A}
	& =
	\sum_{n=0}^{\infty} 
	\begin{pmatrix}
		2 U
	&	0
	\\	0
	&	U
	\end{pmatrix}^n
	=
	\begin{pmatrix}
		( 1 - 2 U )^{-1}
	&	0
	\\	0
	&	( 1 - U )^{-1}
	\end{pmatrix}
, \\
\hat{B}
	& = 
	\begin{pmatrix}
		0
	&	V
	\\	- V
	&	0
	\end{pmatrix}
.\end{align}
 Hence, the VCs up to $\mathcal{O} (n_{\rm i})$ are evaluated as
\begin{subequations}
\begin{align}
\Lambda_{1,x}
	& = \frac{1}{1 - 2 U} \rho_1 \sigma^x
		+ \frac{V}{(1 - U) (1 - 2 U)} \rho_2 \sigma^x
, \\
\Lambda_{2,x}
	& = - \frac{V}{(1 - U) (1 - 2 U)} \rho_1 \sigma^x
		+ \frac{1}{1 - U} \rho_2 \sigma^x
.\end{align}%
\label{eq:L1x_L2x}
\end{subequations}
 Similarly, by interchanging $\mathrm{R}$ and $\mathrm{A}$ in Eqs.~(\ref{eqs:GR_rho1_GA-GR_rho2_GA}), we obtain $\Lambda_{1,x}^{*}$ and $\Lambda_{2,x}^{*}$ up to $\mathcal{O} (n_{\rm i})$ as
\begin{subequations}
\begin{align}
\Lambda_{1,x}^{*}
	& = \frac{1}{1 - 2 U} \rho_1 \sigma^x
		- \frac{V}{(1 - U) (1 - 2 U)} \rho_2 \sigma^x
, \\
\Lambda_{2,x}^{*}
	& = + \frac{V}{(1 - U) (1 - 2 U)} \rho_1 \sigma^x
		+ \frac{1}{1 - U} \rho_2 \sigma^x
,\end{align}%
\end{subequations}
 Note that only the sign of $V$ is different from Eqs.~(\ref{eq:L1x_L2x}).

 By using the above relations and Eqs.~(\ref{eqs:GR_rho1_GA-GR_rho2_GA}), the terms with VCs are obtained as $K_{\mathrm{surf}}^{\mathrm{VC}} = K_{\mathrm{surf}}^{\mathrm{ld}} + K_{\mathrm{surf}}^{\mathrm{sk}}$ with
\begin{align}
K_{\mathrm{surf}}^{\mathrm{ld}}
	& = - \int_{-\infty}^{\infty} \mathrm{d} \epsilon \left( - \frac{\partial f_{\mathrm{FD}}}{\partial \epsilon} \right)	
		\frac{ \pi \Delta \nu (\epsilon) }{\epsilon^2 + \Delta^2} \left(
			\beta (\epsilon) - 1
		 \right)
		+ \mathcal{O} (\gamma^2)
, \\
K_{\mathrm{surf}}^{\mathrm{sk}}
	& = \frac{1}{n_{\rm i} u} \int_{-\infty}^{\infty} \mathrm{d} \epsilon \left( - \frac{\partial f_{\mathrm{FD}}}{\partial \epsilon} \right)
		\frac{\pi \Delta \nu (\epsilon)}{\epsilon} \beta (\epsilon) (U (\epsilon))^2
		+ \mathcal{O} (n_{\rm i}^0 u)
.\end{align}
 It should be noted that $K_{\mathrm{surf}}^{\mathrm{b+ld}} = K_{\mathrm{surf}}^{\mathrm{b}} + K_{\mathrm{surf}}^{\mathrm{ld}}$ is simply expressed as
\begin{align}
K_{\mathrm{surf}}^{\mathrm{b+ld}}
	& = - \int_{-\infty}^{\infty} \mathrm{d} \epsilon \left( - \frac{\partial f_{\mathrm{FD}}}{\partial \epsilon} \right)	
		\frac{ \pi \Delta \nu (\epsilon)}{\epsilon^2 + \Delta^2} \beta (\epsilon)
		+ \mathcal{O} (\gamma^2)
\label{eq:sHc_l+ld}
,\end{align}
and from Eqs.~(\ref{eq:sHc_bare}) and (\ref{eq:sHc_l+ld}), the effect of side-jump is contained in the factor
\begin{align}
\beta (\epsilon)
	& \equiv \frac{1}{(1 - U (\epsilon)) (1 - 2 U (\epsilon))}
	= \frac{9 (\epsilon^2 + \Delta^2)^2}{ 2 (\epsilon^2 + 2 \Delta^2) (\epsilon^2 + 5 \Delta^2) }
,\end{align}
which has a value between $1 \le \beta (\epsilon) \le 9/2$ for $\Delta \le |\epsilon| \le \infty$.

\section{\label{sec:results_Discussion}Results and Discussion}
 We summarize the analytic results of the spin Hall conductivity at $T = 0$,
\begin{align}
\sigma^{z}_{\mathrm{s}, yx} / \sigma_{\rm s}^{0}
	& =	\tilde{\sigma}_{\mathrm{b+ld}}
		+ \tilde{\sigma}_{\mathrm{sk}}
		+ \tilde{\sigma}^{\rm (int)}_{\mathrm{sea}}
\end{align}
with the dimensionless spin Hall conductivities, 
\begin{subequations}
\begin{align}
\tilde{\sigma}_{\mathrm{b+ld}}
	& = \frac{9 (\tilde{\mu}^2 + 1)}{ (\tilde{\mu}^2 + 2) (\tilde{\mu}^2 + 5) } \tilde{\nu}
\label{eq:sHc_b+ld}
, \\
\tilde{\sigma}_{\mathrm{sk}}
	& = - \frac{\Delta}{n_{\rm i} u}
		\frac{ (\tilde{\mu}^2-1)^2 }{ (\tilde{\mu}^2+2) (\tilde{\mu}^2+5) } \frac{\tilde{\nu}}{\tilde{\mu}}
\label{eq:sHc_skew}
, \\
\tilde{\sigma}^{\rm (int)}_{\mathrm{sea}}
	& = - \frac{ \tilde{\nu} }{\tilde{\mu}^2}
	- \int_{-\infty}^{\tilde{\mu}} \mathrm{d} x
		\frac{\tilde{\nu} (x)}{x^3}
\label{eq:sHc_sea}
,\end{align}
\label{eq:sHc_result}%
\end{subequations}
where $\sigma_{\rm s}^{0} = e g^* \Delta / 4 \pi^2 v = m e v / 2 \pi^2$ is the unit of the spin Hall conductivity, $\tilde{\mu} = \mu / \Delta$ is the chemical potential divided by the gap, $\tilde{\nu} (x) = (2 \pi^2 v^3 / \Delta^2) \nu (\epsilon)$ with $x = \epsilon / \Delta$ is the dimensionless DOS [Eq.~(\ref{eq:DOS})], and $\tilde{\nu} = \tilde{\nu} (\tilde{\mu})$ is that at the Fermi level.
 Here, $\tilde{\sigma}_{\mathrm{b+ld}}$ is contributed from the Fermi-surface term of the bare bubble and the ladder type VC, $\tilde{\sigma}_{\mathrm{sk}}$ is a contribution due to the skew scattering, and $\tilde{\sigma}^{\rm (int)}_{\mathrm{sea}}$ is the intrinsic contribution from the Fermi-sea term.

 We now verify that the intrinsic contributions of our results are consistent with those of Fuseya {\it et al}.~\cite{Fuseya2012}.
 Since we have integrated by parts using Eq.~(\ref{eq:DR2-DA2}), $\tilde{\sigma}^{\rm (int)}_{\mathrm{sea}}$ seems to be different from $K^{II}_{syx}$ of Ref.~\cite{Fuseya2012} [Eq.~(10) in Ref. \cite{Fuseya2012}], but $\tilde{\sigma}^{\rm (int)}_{\mathrm{sea}}$ is further calculated as
\begin{align}
\tilde{\sigma}^{\rm (int)}_{\mathrm{sea}}
	= \left\{
		\begin{array}{cc}
		\displaystyle
			\log \left(
				\frac{2 \tilde{E}_{\rm c}}{|\tilde{\mu}| + \sqrt{\tilde{\mu}^2 -1}}
			\right)
		&	( |\tilde{\mu}| > 1 ),
		\\	\displaystyle
			\log ( 2 \tilde{E}_{\rm c} )
		&	( |\tilde{\mu}| \le 1 ),
		\end{array}
	\right.
\end{align}
where we have introduced the energy cut-off, $\tilde{E}_{\rm c} = E_{\rm c} / \Delta e$ such that $\tilde{E}_{\rm c} \gg 1$, following Ref.~\cite{Fuseya2012}.
 The Fermi-surface term of the intrinsic contribution calculated by Fuseya {\it et al}.~\cite{Fuseya2012} can be obtained by dropping $\gamma_3 (\epsilon)$ in Eq.~(\ref{eq:limit_line}) and combining with Eq.~(\ref{eq:sHc_bare}), which reads
\begin{align}
\tilde{\sigma}_{\rm surf}^{\rm (int)}
	& = \frac{\tilde{\nu}}{\tilde{\mu}^2}
	= \frac{ \sqrt{\tilde{\mu}^2 - 1} }{|\tilde{\mu}|} \sum_{\eta = \pm 1} \Theta ( \eta \tilde{\mu} - 1 )
.\end{align}
 In this case, the spin Hall conductivity consists of the contribution just from the states below the Fermi level,
\begin{align}
\tilde{\sigma}^{\rm (int)}_{}
	& = \tilde{\sigma}^{\rm (int)}_{\rm surf} + \tilde{\sigma}^{\rm (int)}_{\rm sea}
	= - \int_{-\infty}^{\tilde{\mu}} \mathrm{d} x
		\frac{ \tilde{\nu} (x)}{x^3}
\label{eq:sHc_intrinsic}
,\end{align}
and there are no Fermi-surface contributions involving $(-\partial f_{\rm FD} / \partial \epsilon)$.
 This feature is shared by the intrinsic anomalous Hall effect~(AHE) due to SOC, which can be described by the Berry phase in momentum space.

\begin{figure}[htbp]
\centering
\includegraphics[width=0.8\linewidth]{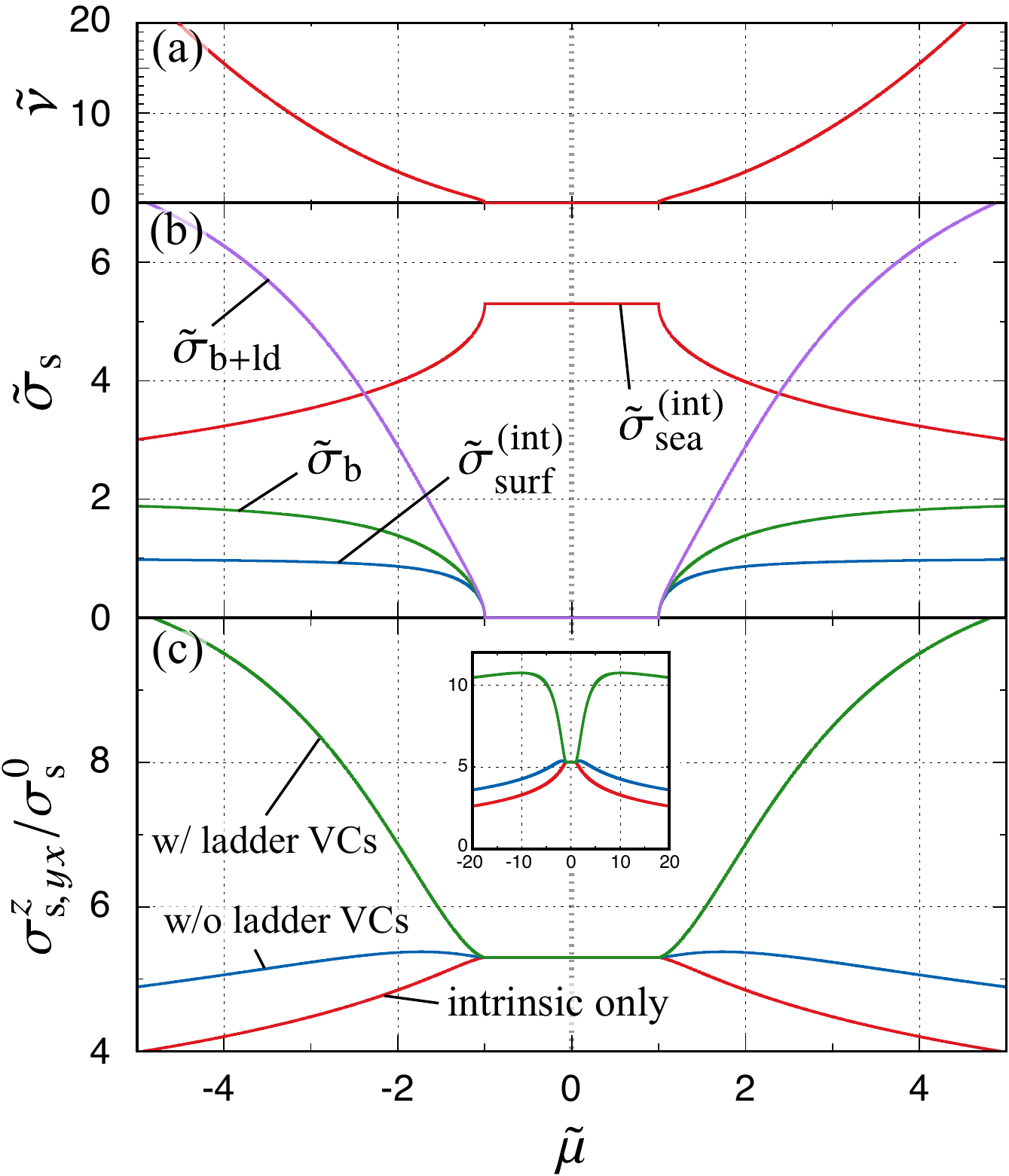}
\caption{\label{fig:DOS+b+l+s}(Color online)~%
 Chemical potential dependences of (a)~the DOS, (b)~the individual $\mathcal{O} (n_{\rm i}^0)$ contributions to the spin Hall conductivity, and (c)~the total spin Hall conductivities without and with the ladder type VCs in addition to the intrinsic one.
 Here, we set $\tilde{E}_{\mathrm{c}} = 100$ following Ref.~\cite{Fuseya2012}.
 The inset of (c) shows a wider region of $\tilde{\mu}$.
 Note that the skew-scattering contribution is not included here, which is $\mathcal{O} (n_{\rm i}^{-1})$.
}
\end{figure}
 Figure~\ref{fig:DOS+b+l+s} shows $\tilde{\mu}$-dependences of DOS and the $\mathcal{O} (n_{\rm i}^0)$ contributions to the spin Hall conductivity.
 The Fermi-surface contributions [Eqs.~(\ref{eq:sHc_b+ld}) and (\ref{eq:sHc_skew})] and the first term of the Fermi-sea term [Eq.~(\ref{eq:sHc_sea})] are proportional to the DOS.
 Hence, these terms vanish when the chemical potential lies in the band gap ($|\tilde{\mu}| < 1$).
 On the other hand, the second term of the Fermi-sea term and the intrinsic contribution [Eq.~(\ref{eq:sHc_intrinsic})] are expressed as the $\epsilon$-integrals of the DOS divided by $\epsilon^3$, and these are finite even in the band gap (Fig.~\ref{fig:DOS+b+l+s} (b)).
 We find that as the extrinsic contributions are taken into account, the total spin Hall conductivity increases, and the extrinsic contributions dominate the intrinsic ones, when the chemical potential is in the band (Fig.~\ref{fig:DOS+b+l+s} (c)).
 Note that when $|\tilde{\mu}| \gg 1$, the spin Hall conductivity with the ladder type VCs is expressed as $\tilde{\sigma}_{\mathrm{b+ld}} + \tilde{\sigma}^{\mathrm{(int)}}_{\mathrm{sea}} \simeq 9 + \log (\tilde{E}_{\mathrm{c}} / |\tilde{\mu}|)$, and thus it changes non-monotonically as seen in the inset of Fig.~\ref{fig:DOS+b+l+s} (c).
 This non-monotonic behavior is because of the damping constants, $\gamma_0$ and $\gamma_3$, also proportional to the DOS, $\tilde{\nu} \propto \tilde{\mu}^2$ for $|\tilde{\mu}| \gg 1$~\cite{footnote3}.

\begin{figure}[htbp]
\centering
\includegraphics[width=0.8\linewidth]{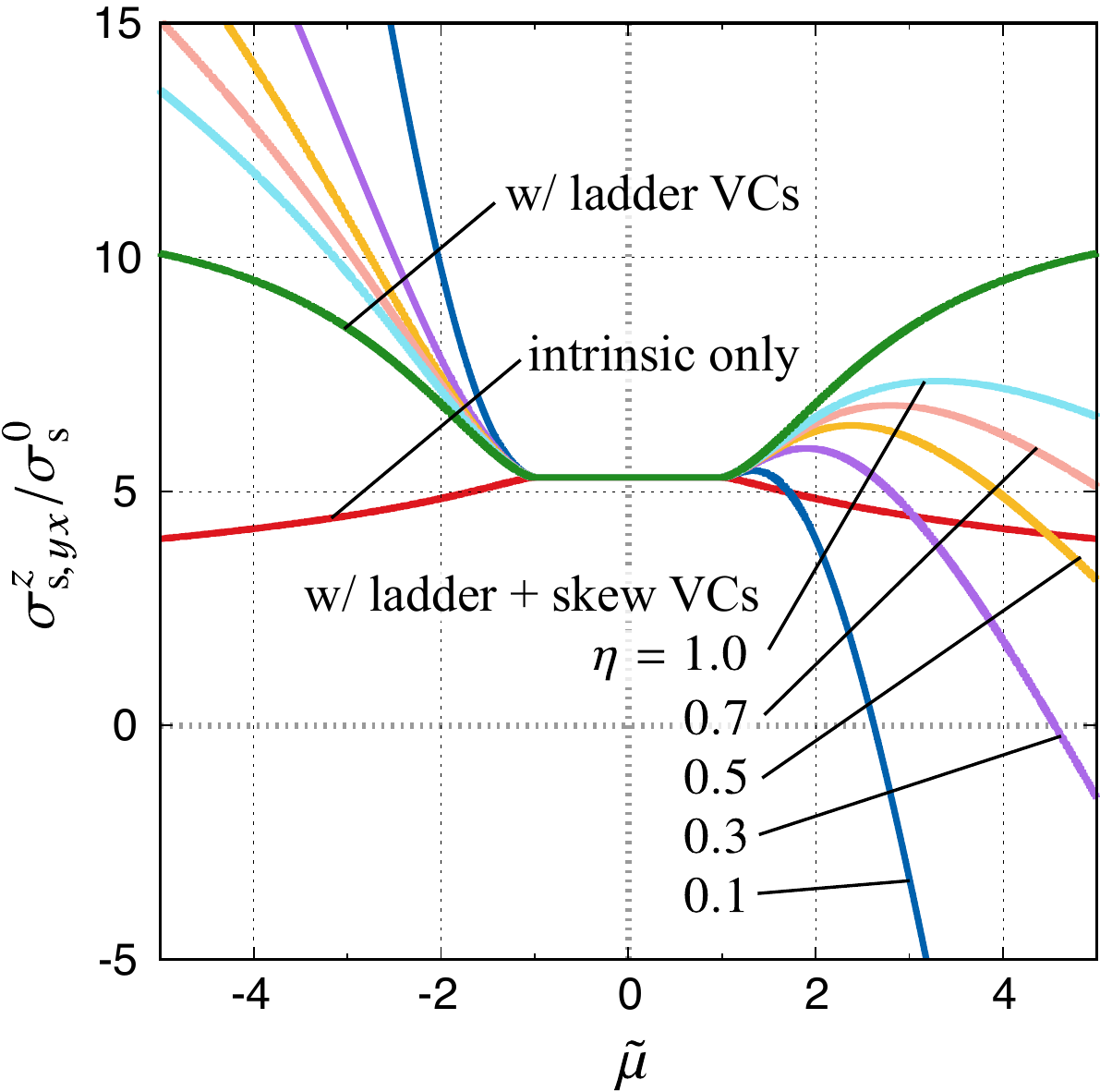}
\caption{\label{fig:total-skew}(Color online)~%
 Chemical potential dependence of the total spin Hall conductivity for $\eta = 0.1$, $0.3$, $0.5$, $0.7$ and $1$, in addition to the intrinsic one and that with the ladder type VCs.
}
\end{figure}
 In Fig.~\ref{fig:total-skew}, the $\tilde{\mu}$-dependence of the total spin Hall conductivity is plotted for values of $\eta = n_{\rm i} u / \Delta$, in addition to the intrinsic spin Hall conductivity and that with the ladder type VCs.
 The skew-scattering contribution is dominant in the clean limit, $\eta \to 0$.
 Particularly for $u > 0$, it is negative for $\tilde{\mu} > 1$ [Eq.~(\ref{eq:sHc_skew})], while the $\mathcal{O} (n_{\rm i}^{0})$ contributions are positive.
 Experiments on the amorphous bismuth~\cite{Emoto2014} and on the polycrystalline bismuth~\cite{Emoto2016} show that the spin Hall angle is not so large as that in platinum, even though the SOC of bismuth is twice as large as that of platinum.
 The authors of Ref.~\cite{Emoto2016} discussed that this discrepancy may arise from extrinsic mechanisms.
 The total spin Hall conductivity for $\tilde{\mu} \simeq 4.5$~\cite{footnote4} is almost zero when $\eta = 0.3$~(Fig.~\ref{fig:total-skew}).
 Our results suggest theoretically the importance of the skew-scattering contribution.
 However, it would be premature to conclude from this that the extrinsic contribution is dominant in realistic situations.
 In our calculation, we have assumed that the impurity potential is proportional to $\rho_0$ for simplicity.
 A different type of the impurity potential $\propto (\rho_0 + \rho_1)$ was proposed~\cite{Sakai1981} based on the $\bm{k}\cdot\bm{p}$ theory.
 In this case, $\rho_3$-component of the self-energy [Eq.~(\ref{eq:damping-constants})] does not exist, and as a result, the skew-scattering contribution is always zero.
 Moreover, there are holes near the $T$-point, which will also contribute to SHE.
 For a more accurate comparison to the experiments, these issues may also deserve further investigation.

 The spin Hall conductivity including the skew-scattering contribution is not simply an even or odd function of $\mu$, as seen in Fig.~\ref{fig:total-skew}.
 It is true that the Dirac Hamiltonian~(\ref{eq:Dirac_Hamiltonian}) is invariant under the particle-hole transformation ($\psi \to \rho_1 \sigma^y \psi^*$ where $\psi$ is a Dirac spinor), but the impurity potential~(\ref{eq:impurity_Hamiltonian}) changes sign and breaks the particle-hole symmetry (so does the chemical potential term). 
 The particle-hole symmetry of the Dirac Hamiltonian~(\ref{eq:Dirac_Hamiltonian}) manifests itself in a way that any physical quantity is invariant under the simultaneous transformation, $\mu \to -\mu$ and $u \to -u$.
 Hence, $\tilde{\sigma}_{\rm b+ld}$ and $\tilde{\sigma}_{\rm sea}$, which are even functions of $u$, are even functions of $\mu$, and $\tilde{\sigma}_{\rm sk}$, which is an odd function of $u$, is an odd function of $\mu$.

 In contrast to the 2D Rashba system, where the spin Hall conductivity vanishes as mentioned in Sec.~\ref{sec:introduction}, we obtain the finite result even when the VCs are included.
 This difference may be understood from the fact that the spin-current operator in 2D Rashba system is proportional to the time derivative of the spin operator and the expectation value of the time derivative vanishes in a stationary state~\cite{Chalaev2005,Dimitrova2005} while the spin velocity of the present Dirac electron system [Eq.~(\ref{eq:spin_velocity})] cannot be expressed using such derivatives.

 Finally, we remark about the physical picture of skew scattering. In general, skew scattering arises when the electron is scattered differently depending on its spin.
 This picture does not apply to the present Dirac electron system since there are no spin-dependent scatterings, as seen in Eq.~(\ref{eq:Im_self-energy}).
 We can see that the Dirac electron is scattered by the impurity potential differently depending on its band (or orbital).

 In conclusion, we have calculated the intrinsic and extrinsic contributions to the spin Hall conductivity on an equal footing, for the (effective) Dirac electron system.
 We assumed the $\delta$-function type impurity potential and treated it within the Born approximation for the self energy and within the ladder type and the skew-scattering type VCs for the spin Hall conductivity, respectively.
 We find that the skew-scattering contribution is proportional to $\Delta / n_{\rm i} u$, and can be dominant over the intrinsic one when the system is metallic.
\begin{acknowledgments}
 We would like to thank M.~Shiraishi for informative discussion.
 J.F. would like to thank Y.~Fuseya for valuable information, A.~Shitade for reading this paper and giving a lot of helpful comments, and H.~Kontani and G.~Tatara for encouraging in publishing this work.
 This work was supported by a Grant-in-Aid for Specially Promoted Research (No.~15H05702) and by JSPS KAKENHI Grant Number 25400339.
\end{acknowledgments}


\begin{thebibliography}{99}
\bibitem{Sinova2015} J. Sinova, S. O. Valenzuela, J. Wunderlich, C. H. Back, and T. Jungwirth, Rev. Mod. Phys. {\bf 87}, 1213 (2015).
\bibitem{Nagaosa2010} N. Nagaosa, S. Onoda, A. H. MacDonald, N. P. Ong, and J. Sinova, Rev. Mod. Phys. {\bf 82}, 1539 (2009).
\bibitem{Murakami2003} S. Murakami, N. Nagaosa, and S.-C. Zhang, Science {\bf 301}, 1348 (2003).
\bibitem{Sinova2004} J. Sinova, D. Culcer, Q. Niu, N. A. Sinitsyn, T. Jungwirth, and A. H. MacDonald, Phys. Rev. Lett. {\bf 92}, 126603 (2004).
\bibitem{Berger1970} L. Berger, Phys. Rev. B {\bf 2}, 4559 (1970).
\bibitem{Smit1955} J. Smit, Physica {\bf 21}, 877 (1955).
\bibitem{Smit1958} J. Smit, Physica {\bf 24}, 39 (1958).
\bibitem{Shubnikov1930}  L. Shubnikov and W. J. de Haas, Comm. Phys. Lab. Leiden {\bf 207d}, 35 (1930).
\bibitem{deHaas1930} W. J. de Haas and P. M. van Alphen, Comm. Phys. Lab. Leiden {\bf 212a}, 3 (1930).
\bibitem{Cohen1960} M. H. Cohen and E. I. Blount, Philos. Mag. {\bf 5}, 115 (1960).
\bibitem{Wolff1964} P. A. Wolff, J. Phys. Chem. Solids {\bf 25}, 1057 (1964).
\bibitem{Fuseya2015} Y. Fuseya, M. Ogata, and H. Fukuyama, J. Phys. Soc. Jpn. {\bf 84}, 012001 (2015).
\bibitem{Wehrli1968} L. Wehrli, Phys. Kondens. Mater. {\bf 8}, 87 (1968).
\bibitem{Fukuyama1970} H. Fukuyama and R. Kubo, J. Phys. Soc. Jpn. {\bf 28}, 570 (1970).
\bibitem{Fuseya2012} Y. Fuseya, M. Ogata, and H. Fukuyama, J. Phys. Soc. Jpn. {\bf 81}, 93704 (2012).
\bibitem{Inoue2004} J. Inoue, G. Bauer, and L. Molenkamp, Phys. Rev. B {\bf 70}, 041303 (2004).
\bibitem{Inoue2006} J. Inoue, T. Kato, Y. Ishikawa, H. Itoh, G. Bauer, and L. Molenkamp, Phys. Rev. Lett. {\bf 97}, 46604 (2006).
\bibitem{Kohn1957} W. Kohn and J. M. Luttinger, Phys. Rev. {\bf 108}, 590 (1957).
\bibitem{Mahan2000} G. D. Mahan, {\it Many-Particle Physics} (Kluwer Academic/Plenum Publishers, New York, 2000) 3rd ed.
\bibitem{Chalaev2005} O. Chalaev and D. Loss, Phys. Rev. B {\bf 71}, 245318 (2005).
\bibitem{Dimitrova2005} O. V. Dimitrova, Phys. Rev. B {\bf 71}, 245327 (2005).
\bibitem{footnote3} {%
	This implies that the Born approximation is no longer valid for $\tilde{\mu} \gg 1$ since the longitudinal charge conductivity ($\propto U (\mu)$) for $T = 0$ becomes constant for $\tilde{\mu} \gg 1$ as seen in Eq.~(\ref{eq:longitudinal_conductivity}).
}
\bibitem{Emoto2014} H. Emoto, Y. Ando, E. Shikoh, Y. Fuseya, T. Shinjo, and M. Shiraishi, J. Appl. Phys. {\bf 115}, 17C507 (2014).
\bibitem{Emoto2016} H. Emoto, Y. Ando, G. Eguchi, R. Ohshima, E. Shikoh, Y. Fuseya, T. Shinjo, and M. Shiraishi, Phys. Rev. B {\bf 93}, 174428 (2016).
\bibitem{Smith1964} G. E. Smith, G. A. Baraff, and J. M. Rowell, Phys. Rev. {\bf 135}, A1118 (1964).
\bibitem{Zhu2011} Z. Zhu, B. Fauque, Y. Fuseya, and K. Behnia, Phys. Rev. B {\bf 84}, 115137 (2011).
\bibitem{footnote4} {%
	For pure semimetallic bismuth, the energy gap and the chemical potential are evaluated as $\Delta \simeq 7.7 \mathrm{meV}$ and $\mu \simeq 35 \mathrm{meV}$~\cite{Smith1964,Zhu2011,Fuseya2012}.
}
\bibitem{Sakai1981} K. Sakai, C. Ishii, and H. Fukuyama, J. Phys. Soc. Jpn. {\bf 50}, 3590 (1981)
\end{thebibliography}
\end{document}